# A New Basis of Geoscience:
# Whole-Earth Decompression Dynamics

**J. Marvin Herndon**
**Transdyne Corporation**
**San Diego, CA 92131 USA**

**Abstract:** Neither plate tectonics nor Earth expansion theory is sufficient to provide a basis for understanding geoscience. Each theory is incomplete and possesses problematic elements, but both have served as stepping stones to a more fundamental and inclusive geoscience theory that I call Whole-Earth Decompression Dynamics (WEDD). WEDD begins with and is the consequence of our planet's early formation as a Jupiter-like gas giant and permits deduction of: (1) Earth's internal composition, structure, and highly-reduced oxidation state; (2) Core formation without whole-planet melting; (3) Powerful new internal energy sources - proto-planetary energy of compression and georeactor nuclear fission energy; (4) Georeactor geomagnetic field generation; (5) Mechanism for heat emplacement at the base of the crust resulting in the crustal geothermal gradient; (6) Decompression-driven geodynamics that accounts for the myriad of observations attributed to plate tectonics without requiring physically-impossible mantle convection, and; (7) A mechanism for fold-mountain formation that does not necessarily require plate collision. The latter obviates the necessity to assume supercontinent cycles. Here, I review the principles of Whole-Earth Decompression Dynamics and describe a new underlying basis for geoscience and geology.

## Introduction

In the *Origin of Continents and Oceans*, Wegener, the father of continental drift theory, the forerunner of plate tectonics theory, stated this (Wegener, 1929): "The determination and proof of relative continental displacements, as shown by the previous chapters, have proceeded purely empirically, that is, by means of the totality of geodetic, geophysical, geological, biological and paleoclimatic data, but without making any assumptions about the origin of these processes. This is the inductive method, one which the natural sciences are forced to employ in the vast majority of cases. The formulation of the laws of falling bodies and of the planetary orbits was first determined purely inductively, by observation; only then did Newton appear and show how to derive these laws deductively from the one formula of universal gravitation. This is the normal scientific procedure, repeated time and again. The Newton of drift theory has not yet appeared."

One might now make similar statements about plate tectonics theory and about Earth expansion theory. Those two theories developed more-or-less simultaneously from entirely different perspectives. Each appears to offer explanations for some geological observations, but each is an incomplete theory sometimes obfuscated by mechanisms and explanations that are either physically impossible or are unsupported by the laws of physics as presently understood. Moreover, neither addresses Earth's formation nor its internal composition. Yet each has helped to provide a path for understanding that ultimately led to a more fundamental and inclusive theory about our planet that I call *Whole-Earth Decompression Dynamics (WEDD).*

Whole-Earth Decompression Dynamics begins with and is the consequence of our planet's early formation as a Jupiter-like gas giant and permits deduction of: (1) Earth's internal composition, structure, and highly-reduced oxidation state; (2) Core formation without whole-planet melting; (3) Powerful new internal energy sources - proto-planetary energy of compression and georeactor nuclear fission energy; (4) Georeactor geomagnetic field generation; (5) Mechanism for heat emplacement at the base of the crust resulting in the crustal geothermal gradient; (6) Decompression-driven geodynamics that accounts for the myriad of observations attributed to plate tectonics without requiring physically-impossible mantle convection, and; (7) A mechanism for fold-mountain formation that does not necessarily require plate collision. The latter obviates the necessity to assume supercontinent cycles. I have described the details and implications of Whole-Earth Decompression Dynamics in a number of scientific articles (Herndon, 2005, 2006c, 2006a, 2007b, 2009a, 2010, 2011b, 2011a, 2012c) and books (Herndon, 2008, 2012b, 2012a, 2012d, 2012e); here I review WEDD principles and describe a new basis for geoscience understanding.

## Background

In 1897, Wiechert (1897) suggested that our planet's bulk density (Cavendish, 1798) might be accounted for if the Earth possesses a core of iron metal similar to iron meteorites. In 1906, Oldham (1906) found that deep within the Earth, seismic waves abruptly enter a low-velocity zone; he had discovered the core. During the next three decades, the core was found to be liquid and its boundary was precisely located at a radius of 3400 km. Then Lehmann (1936) discovered Earth's inner core and correctly estimated its radius to be 1200 km. But next came the question: What is the inner core's chemical composition?

Investigations of earthquake waves, coupled with moment of inertia considerations, can provide information about the existence of different structures within the Earth and whether they are solid or liquid, but not their chemical compositions, which must be inferred from chondrite meteorites. The rock-forming elements of chondrites (like corresponding elements in the photosphere of the Sun) were not appreciably separated from one another and thus provide a basis for understanding the chemical composition of the interior parts of Earth. But a complication arises: There are three groups of chondrites that differ profoundly in their states of oxidation, and thus in their mineral compositions: these groups are designated *ordinary*, *carbonaceous*, and *enstatite* chondrites.

In the late 1930s and early 1940s the idea that the Earth resembles an ordinary chondrite took hold; ordinary chondrites are the most frequent type meteorite seen to fall. Enstatite chondrites were ignored because of their rarity and because their highly-reduced state of oxidation was not understood. So what was the rationale for the chemical composition of the inner core? In the metal of ordinary chondrites, nickel is always alloyed with iron and elements heavier than nickel, even if taken together, would be insufficiently abundant to form the inner core. So, it was assumed that the liquid iron alloy core is in the process of freezing and that the inner core is the frozen iron metal portion.

Later, seismic discontinuities were discovered just above the top of the core, in the mantle at a depth of 660 km, and several more above that depth. Discontinuities, *i.e.*, boundaries where earthquake waves impinging at an angle change speed and direction, can in principle be caused either by (1) chemically different layers or by (2) chemically-identical layers having different crystal structures. Under the assumption that the interior of Earth resembles an ordinary chondrite, (1) was inexplicable, so (2) was assumed.

Four decades after Lehmann discovered the inner core, I realized that discoveries made in the 1960s on enstatite chondrites admit a different possibility for the inner core's composition, namely, fully-crystallized nickel silicide (Herndon, 1979). Subsequently, I demonstrated that that inner 82% of Earth is remarkably similar to a primitive enstatite chondrite by showing in Table 1 that the relative masses of the interior parts of the Earth match corresponding parts of just such a meteorite, the Abee enstatite chondrite (Herndon, 1980, 1993, 2011a). This knowledge is crucial for understanding Earth's origin.

## Origin of Earth as a Jupiter-like Gas Giant

Planetary formation concepts generally fall into one of two categories: (1) *Proto-planetary* concepts that typically involves condensation at high-pressures, hundreds to thousands of atmospheres; or (2) *Planetesimal* concepts that typically involves condensation at very low-pressures, much less than one atmosphere.

Following the 1963 publication of a low-pressure planetary formation model by Cameron (1963), the geoscience community wrongly concurred Earth formed from primordial matter that condensed at low-pressure, one ten-thousandth of an atmosphere (Grossman, 1972) with the dust gathering into progressively more massive rocks, then boulders, planetesimals, and finally planets (Goldrich and Ward, 1973). This composite became accepted as the "standard model of solar system formation". But, I discovered, such low-pressure condensation would lead to terrestrial planets with insufficiently massive cores, as iron would tend to form iron-oxide rather than iron metal (Herndon, 1978, 2006c).

Thermodynamic considerations led Eucken (1944) to conceive of Earth formation from within a giant, gaseous proto-planet when molten-iron rained-out to form the core, followed by the condensation of the silicate-rock mantle. By similar, extended calculations I verified Eucken's results and deduced that oxygen-starved, highly-reduced matter characteristic of enstatite chondrites and, by inference, also the Earth's interior condensed at high temperatures and high pressures from primordial Solar System gas under circumstances that isolated the condensate from further reaction with the gas at low temperatures (Herndon, 2006c; Herndon and Suess 1976).

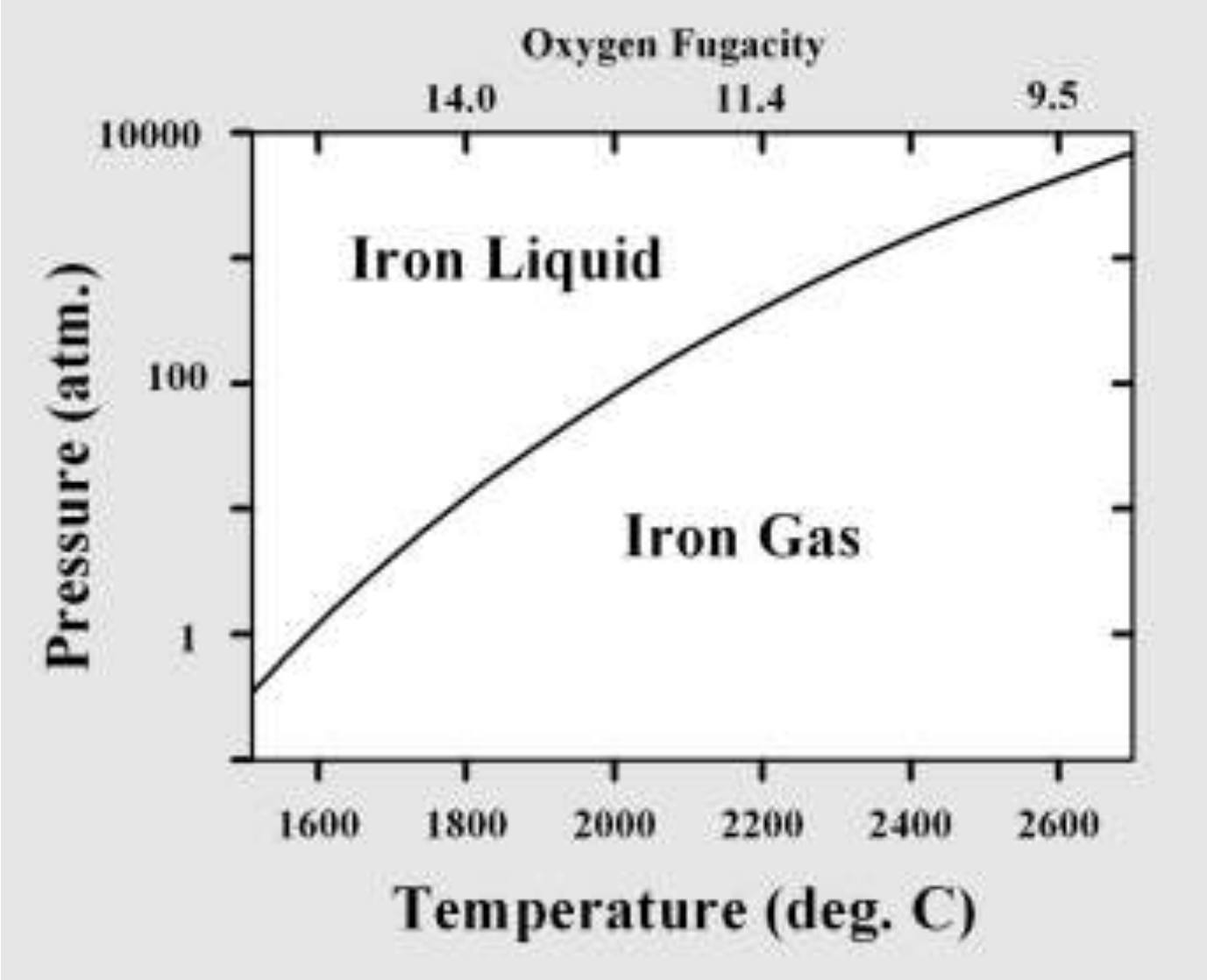

**Figure 1.** The curve in this figure shows the temperatures and total pressures in a cooling atmosphere of solar composition at which liquid iron will ideally begin to condense. The pressure-independent oxygen fugacity is shown on the upper abscissa.

In a cooling gas of solar composition at high pressures and high temperatures, molten iron is the most refractory condensate. **Figure 1** shows the phase boundary between liquid iron metal and iron gas in an atmosphere of solar composition. Ideally, when the partial pressure of a particular substance in the gas exceeds the vapor pressure of that condensed substance, the substance will begin to condense. In a gas of solar composition, the partial pressure of a substance is directly proportional to the total gas pressure, so at higher pressures substances condense at higher temperatures. The degree of oxidation of the condensate, on the other hand, is determined by the gas phase reaction,

$$\mathbf{H_2 + \frac{1}{2}O_2 \leftrightarrow H_2O}$$

which is a function of temperature, but essentially independent of pressure. As I discovered, that reaction leads to an oxidized condensate at low pressures and consequently low temperatures. At high pressures and consequently high temperatures, that reaction leads to highly-reduced enstatite-chondrite-like condensate, provided the condensate is isolated from further reaction with the gas (Herndon, 2006c; Herndon and Suess, 1976) in complete agreement with Eucken's concept of Earth raining-out from within a giant gaseous proto-planet.

I went a step further than Eucken and realized that Earth's complete condensation from a giant gaseous proto-planet yielded a gas-giant planet with about the same mass as Jupiter, because in primordial solar matter, the gases are about 300 times as massive as the rock-forming elements. This is not a strange idea as close to star gas giants are observed in other planetary systems (Seager and Deming, 2010). But Earth is now devoid of primordial gases. What process in nature is capable of removing 300 Earth-masses of primordial gases?

There is a violent period of activity observed in some young stars called the T-Tauri phase characterized by super-intense "solar wind" and mass ejections. It is thought that this brief period is associated with stellar thermonuclear ignition. **Figure 2** is a Hubble Space Telescope image of such an eruption of the binary X, Z Tauri taken in 2000. I have added the white crescent to indicate the leading edge of the plume five years before from a 1995 image of the same event. In five years the leading edge progressed 130 times the distance from the Sun to Earth. A similar event during the Sun's ignition, I posit, stripped the gases from the inner planets of the Solar System and formed the parent matter of ordinary chondrites in the asteroid belt (Herndon, 2007a).

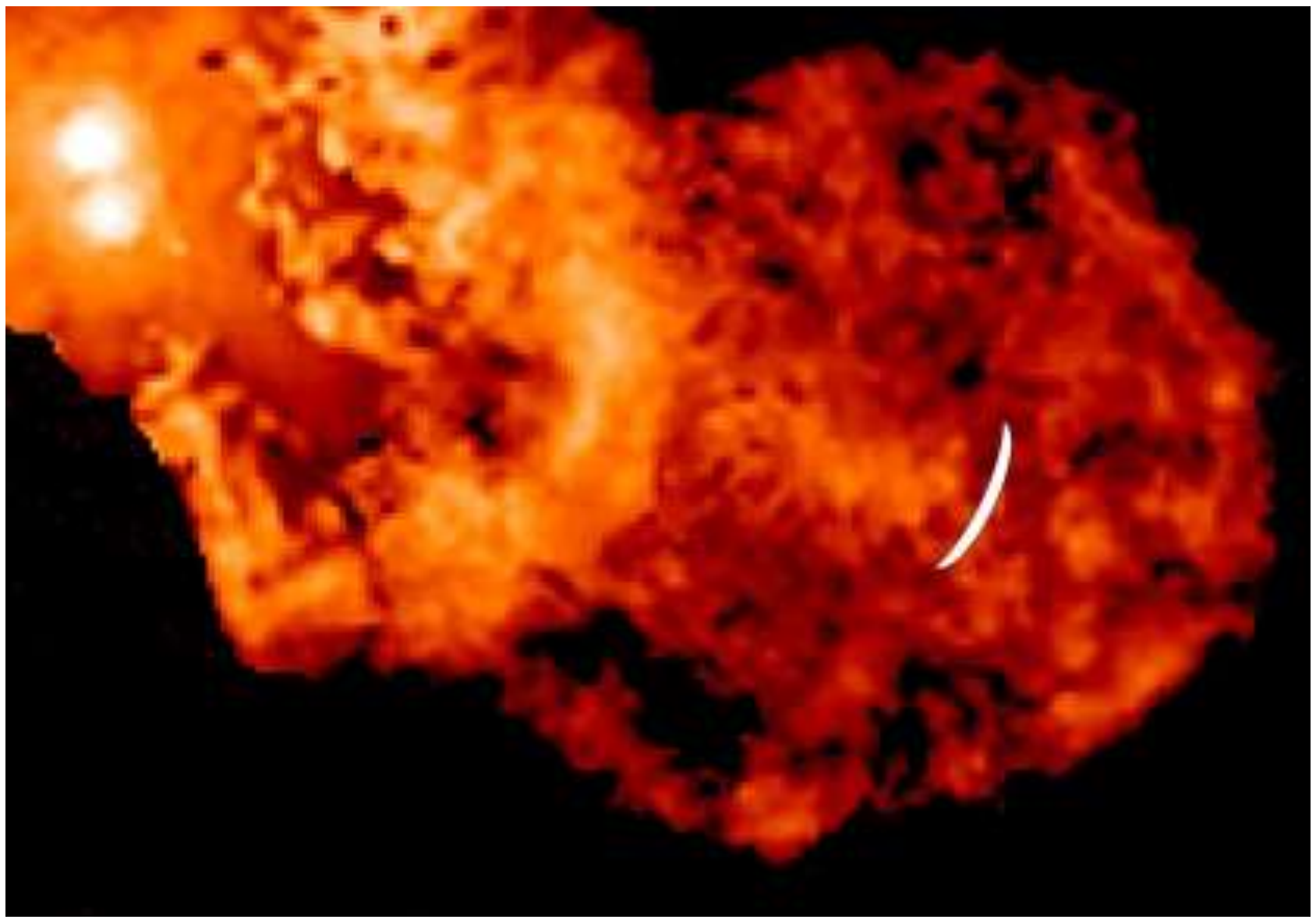

**Figure 2.** Hubble Space Telescope image of binary star X, Z-Tauri in 2000 showing a T-Tauri phase outburst. The white crescent label shows the position of the leading edge of that plume in 1995, indicating a leading-edge advance of 130 A.U. in five years. T-Tauri eruptions are observed in newly formed stars. Such eruptions from out nearly-formed Sun, I submit, stripped the primordial gases from the inner four planets of our Solar System.

Condensation, *i.e.*, raining-out from within a giant gaseous proto-planet at high pressures and high temperatures, led to core formation without whole-planet melting and to a highly reduced internal composition which made possible a planetocentric nuclear fission reactor. Earth's early formation as a Jupiter-like gas giant led to the compression of its rocky kernel and to its subsequent decompression after removal of the massive gaseous overburden. Together, these processes are responsible for Earth's dynamics and crustal geology.

## Consequences of Earth's Condensation

What might one expect as consequences of Earth's formation by condensing, *i.e.*, raining-out from within a giant gaseous proto-planet in the range of pressures and temperatures indicated in **Figure 1**? In this P-T range in an atmosphere of solar composition, thermodynamic calculations indicate that molten iron is less volatile (more refractory) than silicates (Herndon, 2006c; Herndon and Suess, 1976). Thus, as discovered by Eucken (1944), Earth's iron alloy core formed before the mantle silicates. The chemical composition of Earth's core was established during condensation; elements, including some with a high affinity for oxygen (*e.g.*, Ca, Mg, Si), were able to dissolve to some extent in molten iron under those highly-reducing conditions. The mass ratio identities between deep-Earth components and enstatite components, shown in Table 1, means that, to a good approximation, the relative elemental composition of the Earth's deep-interior can be estimated from measurements of the corresponding portion of a primitive enstatite chondrite such as the Abee meteorite (**Table 1**, **Figure 3**). The trace element composition of the core can be similarly estimated because trace elements are slaves to the buffer assemblage formed by the major elements.

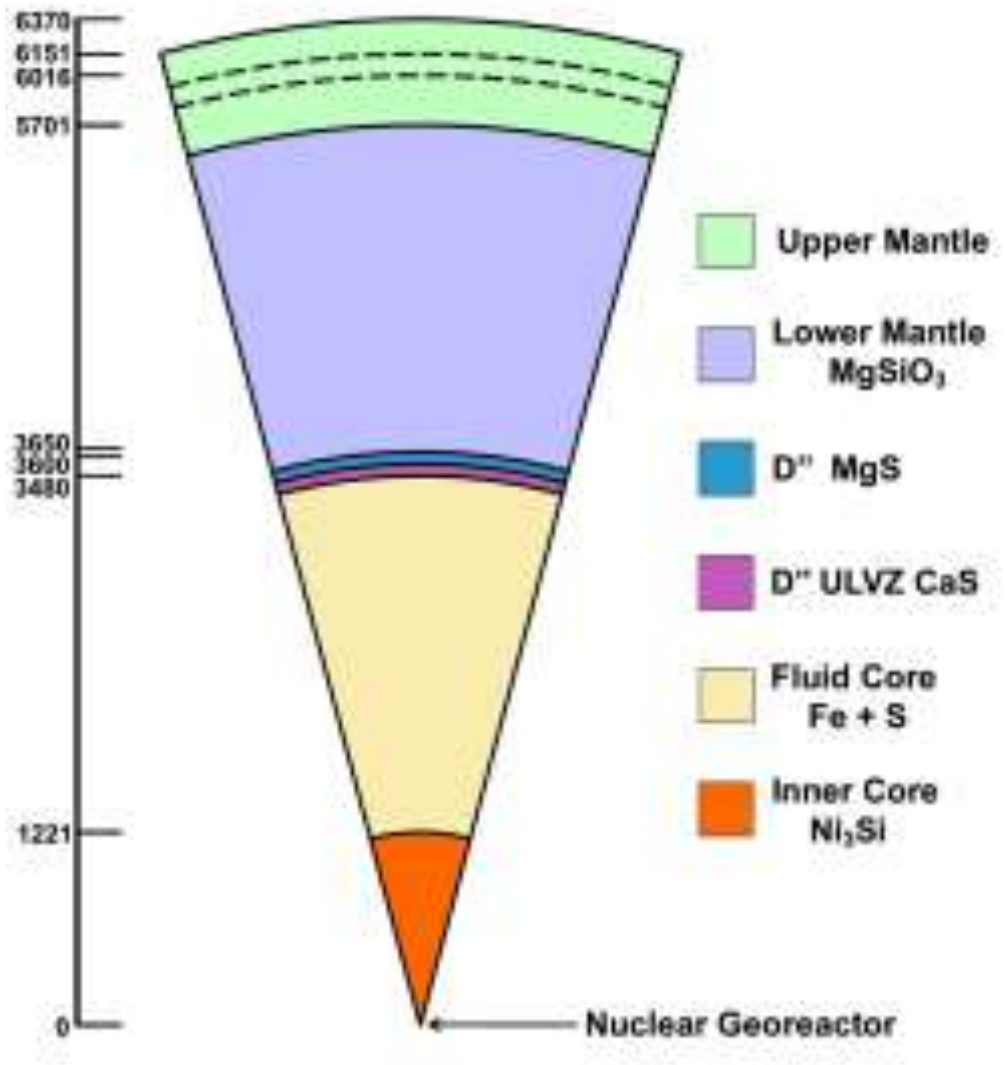

**Figure 3.** Chemical compositions of the major parts of the Earth, inferred from the Abee enstatite chondrite (see **Table 1**). The upper mantle, above the lower mantle, has seismically-resolved layers whose chemical compositions are not yet known. Radius scale in km.

**Table 1.** Fundamental mass ratio comparison between the endo-Earth (lower mantle plus core) and the Abee enstatite chondrite. Above a depth of 660 km seismic data indicate layers suggestive of veneer, possibly formed by the late addition of more oxidized chondrite and cometary matter, whose compositions cannot be specified with certainty at this time.

| **Fundamental Earth Ratio** | **Earth Ratio Value** | **Abee Ratio Value** |
|---|---|---|
| lower mantle mass to total core mass | 1.49 | 1.43 |
| inner core mass to total core mass | 0.052 | theoretical<br>0.052 if $Ni_3Si$<br>0.057 if $Ni_2Si$ |
| inner core mass to lower mantle + total core mass | 0.021 | 0.021 |
| D″ mass to total core mass | 0.09*** | 0.11* |
| ULVZ** of D″ CaS mass to total core mass | 0.012**** | 0.012* |

* = avg. of Abee, Indarch, and Adhi-Kot enstatite chondrites
D″ is the "seismically rough" region between the fluid core and lower mantle
** ULVZ is the "Ultra Low Velocity Zone" of D″
*** calculated assuming average thickness of 200 km
**** calculated assuming average thickness of 28 km
data from (Dziewonski and Anderson, 1981; Keil 1968; Kennet, Engdahl, and Buland, 1995)

In the Abee enstatite chondrite, uranium (also thorium) occurs almost exclusively in the portion corresponding to the Earth's core (Murrell and Burnett 1982). That suggested to me the possibility that the uranium precipitated at some high temperature and, being densest, eventually accumulated at the center of the Earth where it proceeded to function as a natural nuclear fission georeactor (Herndon, 1993, 1994, 1996, 2003; Hollenbach and Herndon, 2001).

The georeactor is thought to consist of a two-part assemblage, as illustrated in **Figure 4**, consisting of a fissioning nuclear sub-core surrounded by a sub-shell of radioactive waste products, presumably a liquid or slurry. The ~24 km diameter assemblage is too small to be presently resolved from seismic data. Oceanic basalt helium data, however, provide strong evidence for the georeactor's existence (Herndon, 2003; Rao, 2002) and antineutrino measurements have not refuted that (Bellini and al., 2010; Gando et al., 2011). To date, detectors at Kamioka, Japan and at Gran Sasso, Italy have detected antineutrinos coming from within the Earth. After years of data-taking, an upper limit on the georeactor nuclear fission contribution was determined to be either 26% (Kamioka, Japan) (Gando et al., 2011) or 15% (Gran Sasso, Italy) (Bellini and al., 2010) of the total energy output of uranium and thorium, estimated from deep-Earth antineutrino measurements (**Table 2**). The actual total georeactor contribution may be somewhat greater, though, as some georeactor energy comes from natural decay as well as from nuclear fission.

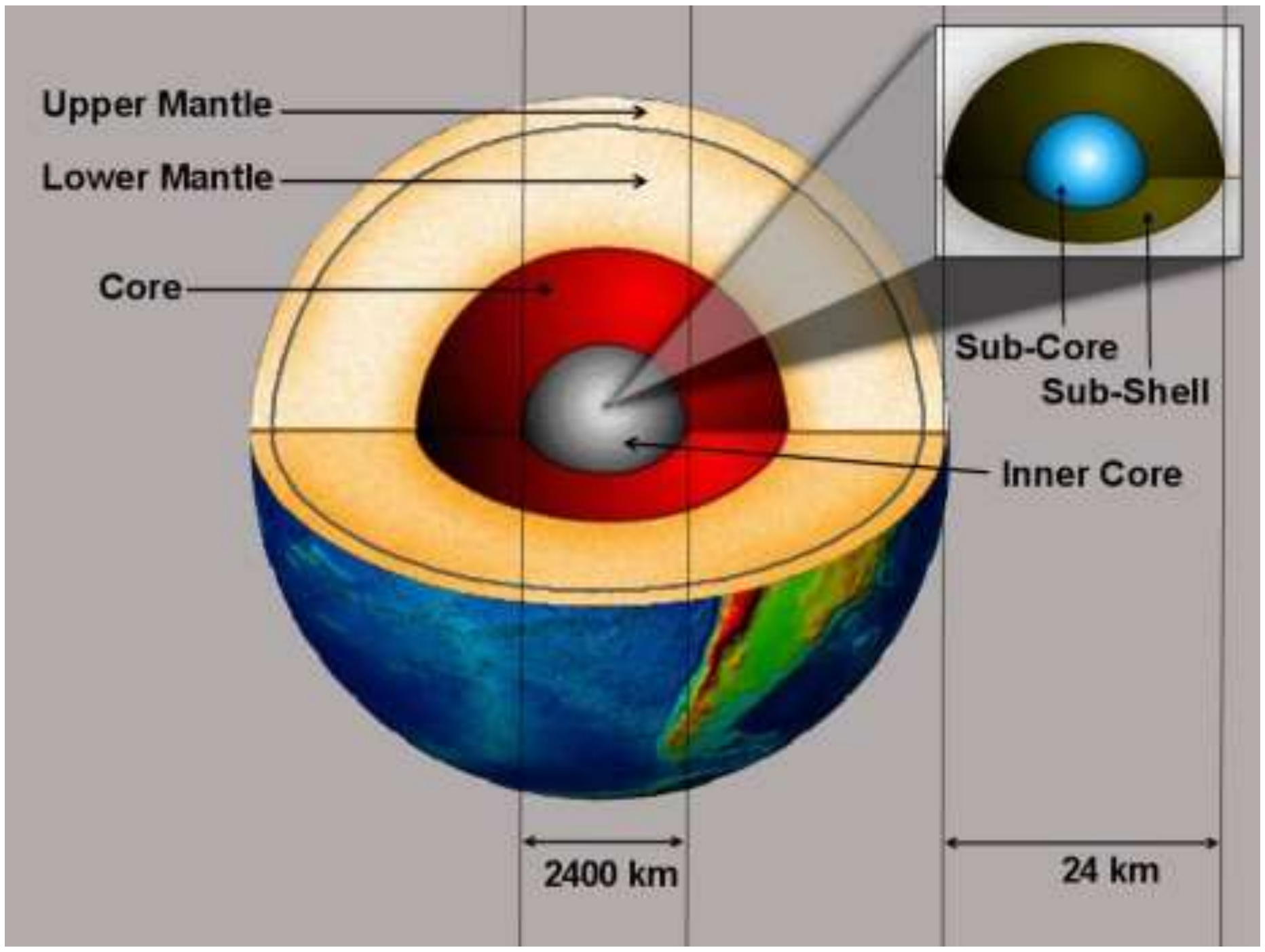

**Figure 4.** Earth's nuclear fission georeactor (inset) shown in relation to the major parts of Earth. The georeactor at the center is one ten-millionth the mass of Earth's fluid core. The georeactor sub-shell, I posit, is a liquid or a slurry and is situated between the nuclear-fission heat source and inner-core heat sink, assuring stable convection, necessary for sustained geomagnetic field production by convection-driven dynamo action in the georeactor sub-shell (Herndon, 1996, 2007b, 2009a).

**Table 2.** Geoneutrino (antineutrino) determinations of radiogenic heat production (Bellini and al., 2010; Gando et al., 2011) shown for comparison with Earth's heat loss to space (Pollack, Hurter, and Johnson, 1993). See original report for discussion and error estimates.

| Heat (terawatts) | Source |
|---|---|
| 44.2 TW | global heat loss to space |
| 20.0 TW | neutrino contribution from $^{238}U$, $^{232}Th$, and georeactor fission |
| 5.2 TW | georeactor KamLAND data |
| 3.0 TW | georeactor Borexino data |
| 4.0 TW | $^{40}K$ theoretical |
| 20.2 TW | loss to space minus radiogenic |

Uranium at the center of Earth and its concomitant functioning as a nuclear fission reactor is a direct consequence of Earth's formation by condensing, *i.e.*, raining-out from within a giant gaseous proto-planet in the highly-reducing environment at high pressures and high temperatures. These further consequences follow: (1) The georeactor provides the energy source and production mechanism for generating the geomagnetic field; (2) Changes in charged particle flux from the Sun, through the intermediary of the geomagnetic field, may induce electric currents into the georeactor that cause ohmic heating and which in extreme instances may cause geomagnetic reversals or excursions; (3) The georeactor provides heat, channeled to the surface, that powers hotspots such as underlie Iceland and the Hawaiian Islands; and, (4) Georeactor heat may augment Earth's decompression.

Elsasser (1939, 1946, 1950) suggested that the geomagnetic field is generated by a convection-driven dynamo mechanism operating in the Earth's fluid core. The dynamo mechanism makes sense for geomagnetic field generation, but, as I discovered, not within the Earth's fluid core. There are periods in geological history when the geomagnetic field has been stable for millions of years. Such long-term convection stability cannot be expected in the Earth's core. Not only is the core-bottom about 23% denser than the core-top, but the core is surrounded by a thermally insulating blanket, the mantle, with lower thermal conductivity and lower heat capacity than the core. Heat brought to the top of the core cannot be efficiently removed; the core-top cannot be maintained at a lower temperature than the core-bottom as necessary for thermal convection. But there is a location near the Earth's center where in principle sustained thermal convection is possible.

I have suggested that the geomagnetic field is produced by Elsasser's convection-driven dynamo operating within the georeactor's radioactive waste sub-shell (Herndon, 2007b). Unlike the Earth's core, sustained convection appears to be quite feasible in the georeactor sub-shell. The top of the georeactor sub-shell is in contact with the inner core, a massive heat sink, which is in contact with the fluid core, another massive heat sink. Heat brought from the nuclear sub-core to the top of the georeactor sub-shell by convection is efficiently removed by these massive heat sinks thus maintaining the sub-shell adverse temperature gradient. Moreover, the sub-shell is not bottom heavy. Unlike in the fluid core, decay of neutron-rich radioactive waste in the sub-shell provides electrons that might provide the seed magnetic fields for amplification.

Helium, trapped in volcanic lava, is observed is a variety of geological settings. The $^{3}He/^{4}He$ ratios measured in basalt extruded at the mid-ocean ridges are remarkably constant, averaging 8.6 times the same ratio measured in air. The $^{3}He/^{4}He$ ratios measured in lava from 18 hot-spots around the globe, such as the Hawaiian Islands, are greater than 10 times the value in air.

Numerical simulations of georeactor operation, conducted at Oak Ridge National laboratory, provided compelling evidence for georeactor existence: Georeactor helium fission products matched quite precisely the $^{3}He/^{4}He$ ratios observed in oceanic basalt as shown in **Figure 5**. Note in that figure the progressive rise in $^{3}He/^{4}He$ ratios over time as uranium fuel is consumed by nuclear fission and radioactive decay. Thermal structures, sometimes called mantle plumes, beneath the Hawaiian Islands and Iceland, two high $^{3}He/^{4}He$ hot-spots, as imaged by seismic tomography (Bijwaard and Spakman, 1999; Nataf 2000) extend to the interface of the core and lower mantle, further reinforcing their georeactor-heat origin. The high $^{3}He/^{4}He$ ratios measured in hot-spot lavas appear to be the signature of "recent" georeactor-produced heat and helium, where "recent" may extend several hundred million years into the past.

The Hawaiian Islands and Iceland are two currently erupting hot-spots with seismic imaging indicating that their heat sources arise from the core-mantle boundary. In each instance the lavas are characterized by high $^{3}He/^{4}He$ ratios, indicative of georeactor-produced heat. Mjelde and Faleide (2009) recently discovered a periodicity and synchronicity through the Cenozoic in lava outpourings from these two hot-spots that Mjelde et al. (2010) suggest may arise from variable georeactor heat-production, which may also begin to explain previously noted correlations between geologic surface phenomena and magnetic reversals (Larson, 1991).

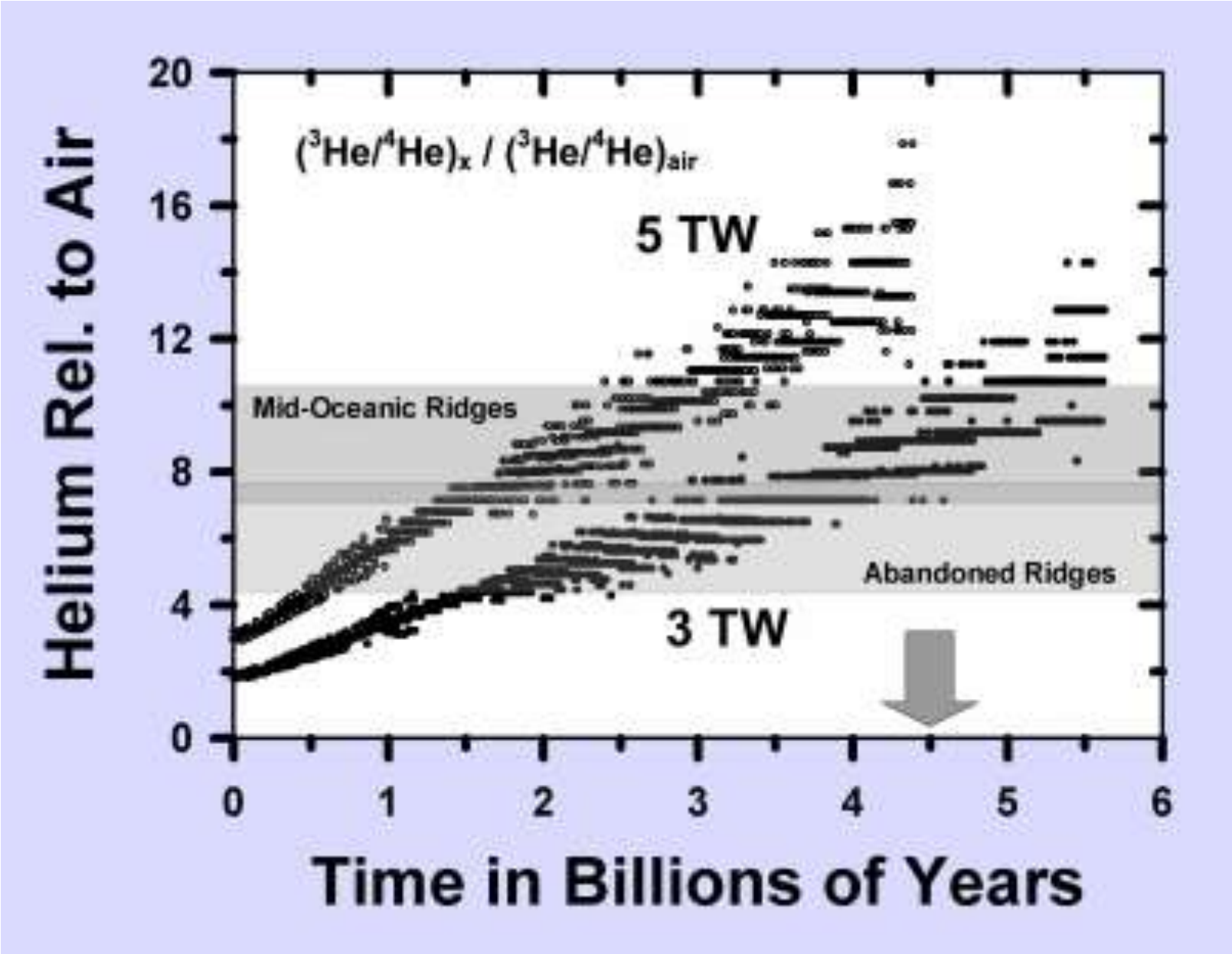

**Figure 5.** Fission product ratio of $^3He/^4He$, relative to that of air, $R_A$, from nuclear georeactor numerical calculations at 5 TW (upper) and 3 TW (lower) power levels (Herndon, 2003). The band comprising the 95% confidence level for measured values from mid-oceanic ridge basalts (MORB) is indicated by the solid lines. The age of the Earth is marked by the arrow. Note the distribution of calculated values at 4.5 Gyr, the approximate age of the Earth. The increasing values are the consequence of uranium fuel burn-up. Iceland deep-source "plume" basalts present values ranging as high as 37 $R_A$ (Hilton et al., 1999).

## Consequences of Earth's Compression

Earth's early formation as a Jupiter-like gas giant has profound consequences on the dynamics of our planet. During Earth's Jupiter-like gas giant stage, the weight of ~300 Earth-masses of gas and ice compressed the rocky kernel to approximately 66% of present diameter. Because of rheology and crustal rigidity, the proto-planetary energy of compression was locked-in when the T-Tauri outbursts stripped away the massive gas/ice layer leaving behind a compressed kernel whose crust consisted entirely of continental rock (sial). Internal pressures began to increase and eventually the crust began to crack.

To accommodate decompression-driven increases in volume in planetary volume, Earth's surface responds in two fundamentally different ways; by crack formation and by the surface folding in response to changes in surface curvature. Cracks form to increase the surface area required as a consequence of planetary-volume increases. *Primary* cracks are underlain by heat sources and are capable of basalt extrusion; *secondary* cracks are those without heat sources that ultimately become repositories for basalt extruded by primary cracks. Decompression-driven increases in volume result in a misfit of the continental rock surface formed earlier at a smaller Earth-diameter. This misfit results in excess surface material confined within continent margins, which adjusts to the new surface curvature by buckling, breaking and falling over upon itself producing fold-mountain chains as illustrated in **Figure 6**. Another less significant adjustment to surface curvature causes peri-continental tension fractures that may initiate the formation submarine canyons and drainage systems (Herndon, 2012c).

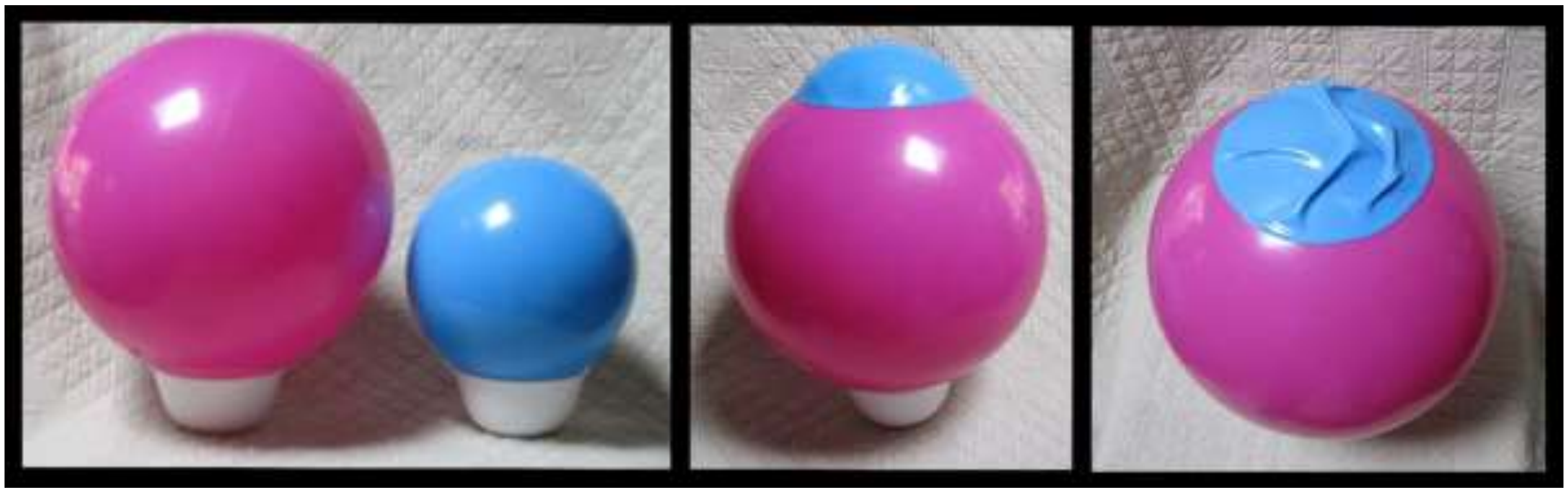

**Figure 6.** Demonstration illustrating the formation of fold-mountains as a consequence of Earth's early formation as a Jupiter-like gas giant. On the left, two balls representing the relative proportions of 'present' Earth (large), and 'ancient' Earth (small) before decompression. In the center, a spherical section, representing a continent, cut from 'ancient' Earth and placed on the 'present' Earth, showing: (1) the curvature of the 'ancient continent' does not match the curvature of the 'present' Earth and (2) the 'ancient continent' has 'extra' surface area confined within its fixed perimeter. On the right, tucks remove 'extra' surface area and illustrate the process of fold-mountain formation that is necessary for the 'ancient' continent to conform to the curvature of the 'present' Earth. Unlike the ball-material, rock is brittle so tucks in the Earth's crust would break and fall over upon themselves producing fold-mountains.

As the Earth decompresses, heat must be supplied to replace the lost heat of proto-planetary compression. Otherwise, decompression would lower the temperature, which would impede the decompression process. Heat generated deep within the Earth from georeactor nuclear fission and natural decay of radionuclides may enhance mantle decompression by replacing the lost heat of proto-planetary compression. The resulting decompression, beginning within the mantle, will tend to propagate throughout the mantle, like a tsunami, until it reaches the impediment posed by the base of the crust. There, crustal rigidity opposes continued decompression; pressure builds and compresses matter at the mantle-crust-interface, resulting in compression heating (Herndon, 2006a). Mantle Decompression Thermal-Tsunami, as I call it, poses a new explanation for heat emplacement at the base of the crust, a new basis of the geothermal gradient, and may be involved in earthquakes and volcanism, as these geodynamic processes appear concentrated along decompression cracks. Moreover, heat emplaced at the base of the crust may be involved in the formation of abiotic hydrocarbons (Herndon, 2006b, 2010). Note that heat emplacement by Mantle Decompression Thermal-Tsunami is of distinct origin from georeactor heat that is channeled from the core to the surface.

## A New Understanding of Geology

The ancient supercontinent, Pangaea, envisioned by Wegener (1912) and adopted in plate tectonic considerations was thought to be surrounded by ocean occupying nearly 1½ times its surface area. By contrast, in Whole-Earth Decompression Dynamics, there is but one true ancient supercontinent, the 100% closed contiguous shell of continental rock covering early Earth's compressed kernel. I call that supercontinent *Ottland* in honor of Ott Christoph Hilgenberg, who first conceived of its existence (Hilgenberg, 1933).

Hilgenberg envisioned a smaller early Earth without ocean basins that had a radius 50% of present day Earth, which he calculated from the total surface area above sea water. I and others calculated a radius of 64% the present day value by considering the surface area of submerged continent margins. Taking into account the folded surface area from mountain formation (Herndon, 2012c), I estimate the early Earth radius was approximately 4200 km, 66% of the present day Earth radius.

The fragmentation of Ottland by decompression cracks and the subsequent formation of interstitial ocean basins neither need to commence immediately upon removal of primordial gases nor necessarily occur at a single point in time. The yet-to-be-delineated sequences of Ottland subdivision and concomitant displacement may bear only superficial resemblance to the popular, but hypothetical, breakup of Pangaea, where the continents are assumed free to wander, breaking up and re-aggregating, while riding atop non-existent mantle convection cells. In Whole-Earth Decompression Dynamics there are fewer degrees of freedom. While complex, there is simplicity as well because continent fragmentation and dispersion represents the approach toward dynamical equilibrium, not the random breaking up and re-aggregating previously imagined.

Whole-Earth Decompression Dynamics extends plate tectonic concepts as it is responsible for Earth's well-documented features without requiring problematic mantle convection (Herndon, 2005, 2010, 2011a). Fold-mountain formation is a natural consequence of whole-Earth decompression without requiring plate collisions (Herndon, 2012c); mountains pushed up from beneath or occurring at colliding plate boundaries, however, are not excluded (Ollier and Pain, 2000). Partially in-filled secondary decompression cracks uniquely explain oceanic troughs, inexplicable by plate tectonics. And, compression heating at the base of the rigid crust is a direct consequence of mantle decompression (Herndon, 2006a). Plate tectonic meanings and terminology are to a great extent preserved in my new paradigm. For example, *rift* is similar, but with different causal mechanism; *transform plate boundaries* are identical; *divergent plate boundaries* are similar, but with a different driving mechanism; *convergent plate boundaries* likewise are similar, but down-plunging plates neither create oceanic trenches, which are secondary decompression cracks, nor are they recycled through the mantle by conveyer-like mantle convection, and; *Wadati-Benioff zones* are quite similar, with the possible exception of why mantle melting occurs that is responsible for sometimes associated volcanic eruptions. Many of the plethora of observations taken to support plate tectonics support as well Whole-Earth Decompression Dynamics.

The processes involved in whole-Earth decompression are yet evident, but progress at a much slower pace than in the geological past. If the Earth is presently decompressing, length of day measurements should show progressive lengthening. Such measurements, made with increasing precision over the last several decades, show virtually no current lengthening (Chao, 1994), implying no significant current decompression crack formation. But major secondary decompression cracks are still conspicuously evident, for example, the circum-pacific trenches, such as the Mariana Trench. And, the Whole-Earth Decompression Dynamics process of basalt extrusion and crack in-filling continues at present. The amount of present decompression, estimated from cumulative basalt outpouring, is on the order of the error limit of length of day measurements. Decompression crack formation might be episodic, though, like the release of stress by major earthquakes.

The Afar triangle is the triple junction where the Red Sea rift, the Carlsberg Ridge of the Indian Ocean, and the East African Rift System meet. Seismic tomographic imaging beneath that region shows a very large, low-velocity zone extending to the base of the lower mantle, referred to as a "superplume" (Ni et al., 2002; Zhao, 2001). Because heat at the base of the lower mantle cannot make bottom-mantle matter sufficiently buoyant to float upward (Herndon, 2009b), the high $^{3}He/^{4}He$ ratios, $R_A$>10, measured in Afar volcanic basalt (Marty et al., 1993), indicate georeactor heat channeling, which allows the highly mobile, inert helium to migrate upward toward regions of progressively-lower density. Throughout geological history georeactor heat channeled from the core appears to be associated with continent fragmentation.

Hotspot formation beneath the continental-rock surface and the localized swelling produced by sub-surface magma augments continent fragmenting that provides increased surface area to accommodate decompression-increased planetary volume. The stress of whole-Earth decompression along with the appearance of an underlying hotspot appears to be the combined-agent involved in continent fragmenting and the concomitant opening of new ocean basins. The georeactor-powered hotspot that underlies the current break-up of Africa along the East African Rift System appears to have contributed to the opening of Gulf of Eden and the Red Sea. Similarly, the opening of the North Atlantic Ocean 61 million years ago appears to be associated with the Iceland hotspot. How universal is the association of continent-fragmentation with georeactor-produced heat? The details are not entirely clear; there is much yet unknown.

The presence of a georeactor-powered hotspot in combination with decompression-stress does not always result in continent splitting. The region located between the Ural Mountains and the Siberian Platform, including the West Siberian Basin, and the Siberian Platform, underwent extensive rifting about 500-250 million years ago (Saunders et al., 2005). Uplifting led to rift-basin formations that developed geological strata on a grand scale, extremely conducive for trapping petroleum and natural gas (Reichow et al., 2002). About 250 million years ago, massive basalt floods, known as the Siberian Traps, spewed forth for about one million years, blanketing the area with perhaps more than 2,000 $km^3$ of basalt containing helium with high $^3He/^4He$ ratios (Basu et al., 1995). Evidence indicates that rifting continued after basalt extrusion (Reichow et al., 2002). Rifting appears to underlie the formation of rift-basins and later the massive eruption of basalt. The high $^3He/^4He$ ratios indicate georeactor heat channeled from the Earth's core. Today, that area is known to contain some of the most extensive petroleum, natural gas, and coal deposits in the world.

For four decades, petroleum exploration geology has been described in terms of plate tectonics, which is based upon physically-impossible mantle convection (Herndon, 2009b). Surface manifestations, such as foreland basins and back-arc basins, which seem well-described by plate tectonics, are in many instances similarly described by Whole-Earth Decompression Dynamics, although with subtle, but importance, differences, especially those related to the absence of convection-driven subduction.

In plate tectonics terminology, "rift" refers to the interface of two plates that are beginning to pull apart. In Whole-Earth Decompression Dynamics, "rift" refers to the beginning of the formation of a decompression crack. Rifting is an integral part of the whole-Earth decompression process of successive continent fragmenting, beginning with Ottland and continuing into the present. The process of continent fragmenting begins with the formation of a decompression crack. Over time, the crack widens, forming a rift-valley or basin. Volcanic eruptions may subsequently occur, depending mainly upon available heat. The rift-basin thus formed becomes an ideal environment for the development of geological strata frequently associated with petroleum and natural gas deposits and can remain a part of the continental margins even after ocean basin formation.

I have suggested that virtually all petroleum and natural gas deposits are connected in some way to, or are the consequence of the WEDD process of rifting, even those deposits, such as foreland basins, that involve underthrust compression, which may result from rifting and extension elsewhere (Herndon, 2010). Continent fragmenting, both successful and failed, initiates with rifting. Observations of rifting which is currently taking place at the Afar triangle in northeastern Ethiopia, and observations of the consequences of rifting throughout the East African Rift System can help to shed light on the nature of petroleum-deposit related rifting that has occurred elsewhere.

Whole-Earth Decompression Dynamics processes observed at Afar and along the East African Rift System provide all of the crucial components for petroleum-deposit formation. Rifting causes the formation of deep basins, as evidenced, for example, by Lake Tanganyika, depth 1.4 km, the second deepest lake in the world, and by Lake Nyasa, depth 0.7 km, the fourth deepest lake, both of which occur as part of the East African Rift System. The observed uplifting caused by swelling from below (Almond, 1986) makes surface land susceptible to erosion, thus providing great amounts of sedimentary material for reservoir rock in-filling of basins. Volcanic-derived sedimentary material may be richly mineralized, assuring strong phytoplankton blooms. Uplifting can sequester sea-flooded lands thus leading to the formation of marine halite deposits through desiccation, and can also lead to dome formation as well.

In the unchanging global-dimension of plate tectonics, the supercontinent Pangaea was thought to be surrounded by ocean. In that view, Pangaea-fragmentation simply shifted land and ocean volumes around without any major change in level. The only mechanism envisioned in that paradigm for rapid, major lowering of sea level is the “ice age” when a large volume of ocean water is sequestered as polar and glacial ice. But 65 million years ago, for example, the time of one major sea level low, the climate was too warm for an ice age. Whole-Earth Decompression Dynamics, by contrast, offers an additional explanation for the lowering of sea level caused by increased planetary diameter and, especially, by continent fragmentation.

The combined-agency of whole-Earth decompression and the delivery of georeactor heat via hotspots is responsible, not only for continent fragmenting with the concomitant opening of new ocean basins, and the origination of petroleum and natural gas deposits, but, I posit, also has had a pronounced impact on life. For example, one of the major spikes of species extinction occurred 65 million years ago at the end of the Cretaceous and beginning of the Tertiary. That is the time of the georeactor-powered massive basalt flood in India that produced the Deccan Traps, which is thought by some to also be associated with the opening of the Arabian Sea. Species extinction at the K-T boundary, as some believe, may be more involved than simply the consequence of an asteroid impact (Courtillot, 2002).

Expected WEDD consequences on species extinction might include these: (1) Massive volcanic eruptions would wreak havoc on the environment on a global scale, as some have stated; (2) Opening of new decompression cracks that form a new ocean basin would lower the existing sea level with devastating impact on some biota; (3) Ocean temperatures might rise due to channeled georeactor-heat and exposure to heat emplaced at the base of the crust by Mantle Decompression Thermal Tsunami; (4) Ocean circulation patterns and prevailing winds would be disrupted, and; (5) The ocean might become acidified, *i.e.*, the pH lowered. The last requires some explanation.

Hard-rock miners, who tunnel into the Earth’s crust to discover and extract valuable ore, first encounter an oxidized zone. Over time oxygen-bearing ground-water percolated through that zone and oxidized some of the minerals found there. Upon tunneling deeper, the miners enter a zone not oxidized in that manner. There they encounter “reduced” minerals such as arsenic-bearing pyrite. Water exposed to this environment, for example, through flooded mine shafts becomes acidified. Residues from mining at this level, *i.e.*, the tailings left on the surface, over time oxidize and release toxic heavy metals into the environment. It would not be surprising that a continent thus split apart might acidify and contaminate ocean water. One might expect similar acidification and contamination from the acid-rain produced by volcanic eruptions of the scale and rapidity of the Deccan Traps. Over time the pH will return to neutral through the reaction of acidic seawater with calcium carbonate.

The primary impetus for the observed spikes in species extinction, I posit, is WEDD continent-fragmenting and the consequent-processes attendant thereto, such as the volcanic activity that causes massive basalt floods. It should not go unnoticed that the greatest spike in mass extinctions occurred about 250 million years ago, contemporaneous with the georeactor-heat powered massive volcanic eruptions that yielded the Siberian Traps at the end of the Permian and beginning of the Triassic. And, there is another "big five" spike in species extinction that occurred about 200 million years ago at the time of the opening of the Central Atlantic Ocean. Mass extinctions are not separate, isolated events but parts of the ongoing process of whole-Earth decompression that is a consequence of Earth's early formation as a Jupiter-like gas giant. Other massive extinctions even further back in time presumably were of similar origin, such as the event some 436 million years ago that marked the end of the Ordovician Period and the beginning of the Silurian. The challenge for geologists is to discover the time series of continent-fragmenting events, successful and failed, that led to the opening of Earth's ocean and lake basins and to the concomitant devastation of prevailing biota.

Within the flawed framework of plate tectonics, explanations were devised to explain the inexplicable. For example, observations of mountain ranges that predate the supposed formation of Pangaea were inexplicable, and so were explained by hypothetical supercontinent cycles, also called Wilson cycles. The underlying idea is that supercontinents would form, and then break-up, and then reform, etc. so that the collisions involved would explain various mountain ranges. I have referred to these as "fictitious supercontinent cycles" (Herndon, 2013); fold-mountain formation is a natural consequence of adjustments to changes in surface curvature resulting from Earth's early origin as a Jupiter-like gas giant (Herndon, 2012c). Indeed, virtually all of geoscience is the consequence of Earth's origin by raining-out from primordial matter at high pressures at high pressures and high temperatures and forming initially as a Jupiter-like gas giant, just as described by Whole-Earth Decompression Dynamics.

I have described the framework of Whole-Earth Decompression Dynamics as an artist might paint: with broad strokes of the brush. Within that framework others may find a richness of detail that will undoubted lead to further advances. In instances, observations are compelling. Secondary decompression cracks are clear and evident: Lake Baikal, for example, or the Mariana Trench. At Afar in Ethiopia and along the East African Rift System the beginnings of continent fragmentation is currently ongoing. But the many details between the initiation of a rift and its subsequent development into a mid-ocean ridge are yet completely unknown; perhaps insights may come from studying ocean floor drill-cores. Because the mass of the georeactor is just one ten-millionth the mass of Earth's core, changes in the charged particle flux from the Sun can, through the intermediary of the geomagnetic field, induce electrical currents into the georeactor causing ohmic heating. In principle such induced heat might disrupt sub-shell convection and lead to magnetic reversals or excursions. Correlations between geologic surface phenomena and magnetic reversals (Larson, 1991) suggest possible connection to large changes in solar charged particle flux. Simultaneity is basalt outpourings on opposite sides of the globe appear to be tied to variations in georeactor heat production (Mjelde and Faleide, 2009; Mjelde, Wessel, Mùller, 2010). And, what about the connection to massive flood basalts? Clearly there is much to be discovered within the framework of Whole-Earth Decompression Dynamics.

## Acknowledgements

This work is dedicated to the memory those whose friendship, inspiration, and encouragement helped to light the way: Paul K. Kuroda, Inge Lehmann, Lynn Margulis, Hans E. Suess, and Harold C. Urey. I have benefited from the comments and/or work of many, notably including Asish R. Basu, Richard B. Cathcart, J. Freeman Gilbert, Rudolph Gottfried, Giovanni Gregori, David R. Hilton, Cliff Ollier, James Maxlow, and Giancarlo Scalera. I thank Dong R. Choi for encouragement.